\documentclass[journal,onecolumn]{IEEEtran}
\usepackage{epsfig}
\usepackage{graphicx}
\usepackage{graphics}
\usepackage{epstopdf} 
\usepackage{xcolor}
\usepackage{pgf, tikz}
\usetikzlibrary{calc,shapes.geometric}
\usetikzlibrary{snakes,backgrounds}
\usetikzlibrary{quotes,angles}
\usetikzlibrary{automata, positioning}
\usetikzlibrary{shapes,decorations,arrows,calc,arrows.meta,fit,positioning}
\tikzset{
    -Latex,auto,node distance =1 cm and 1 cm,semithick,
    state/.style ={ellipse, draw, minimum width = 0.7 cm},
    point/.style = {circle, draw, inner sep=0.04cm,fill,node contents={}},
    bidirected/.style={Latex-Latex,dashed},
    el/.style = {inner sep=2pt, align=left, sloped}
}

\usepackage{amssymb}
\usepackage{multirow}
\usepackage{tcolorbox}% for framed rounded boxes
\tcbset{colframe=black,colback=white,colupper=black,
fonttitle=\bfseries,nobeforeafter,center title,size=small,arc=0pt,outer arc=0pt}

\ifCLASSINFOpdf
\else
\fi

\usepackage[cmex10]{amsmath}

\def\clock{\n=\time \divide\n 60
  \m=-\n \multiply\m 60 \advance\m \time
  \ifnum \n>12 \advance\n -12 \fi
   \number\n.\twodigits\m~\ampm\time}
\def\ampm#1{\ifnum #1< 720 am\else pm\fi}
\def\twodigits#1{\ifnum #1<10 0\fi \number#1}

\def\nexto{\kern -0.54em}
\def\mean{{\rm {I\ \nexto E}}}
\def\prob{{\rm {I\ \nexto P}}}

\def\I{{\rm {I\ \nexto I}}}

\def\U{{\cal U}}

\newcount\m \newcount\n
\def\clock{\n=\time \divide\n 60
  \m=-\n \multiply\m 60 \advance\m \time
  \ifnum \n>12 \advance\n -12 \fi
   \number\n.\twodigits\m~\ampm\time}
\def\ampm#1{\ifnum #1< 720 am\else pm\fi}
\def\twodigits#1{\ifnum #1<10 0\fi \number#1}

\begin{document}

%TITLE AND AUTHOR
\title{Preliminary Model for Operational Effectiveness Assessment of Dismounted Combat Teams}
\author{Graham  V. Weinberg \\ 
Defence Science and Technology Group, P. O. Box 1500, Edinburgh, 5111, South Australia\\
(Draft created at \clock)\\
Graham.Weinberg@defence.gov.au 
 }
\maketitle

% The paper headers
\markboth{Preliminary Model for Operational Effectiveness Assessment of Dismounted Combat Teams: \today}%
{}

\begin{abstract} 
This paper is concerned with introducing a framework in which the operational effectiveness of a dismounted combat team can be assessed. It has been motivated by studies at Defence Science and Technology Group which have focused on the structure of the future dismounted fighting force. In order to determine the operational effectiveness of different combat team designs a metric, based upon simple additive weighting of warfighting functions, is proposed. This then allows a single measure of expected performance to be produced. The model itself is a preliminary attempt to capture essential features of dismounted combat, and the study will indicate how it can then be applied in practice. Discussion on model refinement, based upon subject matter expert data elicitation, is also included.
\end{abstract}

\section{Introduction}
\label{sec:1}
Military forces have experienced significant capability enhancements through the application of technology \cite{ohanlon}, and the land-based fighting force is still undergoing transformations through the application of new technological systems. Hence, since technology is being applied to the fundamental unit of a group of soldiers in the operational environment, it is of importance to quantify the resultant gains to the combat team in terms of achieving its mission success \cite{kott, billing}. Consequently, this study is concerned with the development of a framework in which the operational effectiveness of a dismounted combat team can be assessed.

From an Australian Army perspective, a dismounted combat team consists of a company of soldiers organised into a series of three platoons. Each such platoon contains three sections, where the latter is a group of nine soldiers. Each section has its own equipment selected for the mission being undertaken in an operational setting. The analysis in this paper will be applied at the combat team level, but can be specialised to analysis of different section structures \cite{LWD1}.

Defence Science and Technology Group (DSTG) has been exploring the impact of current and emerging technology on the future dismounted fighting force. Towards this objective, 
the concept of a semi-autonomous combat team (SACT) was developed in recent years \cite{sawers, kempt}. This concept introduces robotic and semi-autonomous systems into the combat team, focusing on their integration and the dynamics of operation in future scearios. In particular, the SACT study explored what a dismounted combat team would look like in the 2030 timeframe. At the section level, a conventional dismounted combat team would be equipped with portable weapons, communications devices and surveillance apparatus including small drones.  By contrast, a SACT section reduces the human element to six soldiers, who have enhanced technology, and the section is 
partitioned into two subsets, one responsible for intelligence, surveillance and reconaissance (ISR) 
operations, and the second for direct engagement with enemy forces. The ISR subsection is equipped with enhanced drones and ground-based robots, while the direct engagement subsection has unmanned ground vehicles (UGVs) equipped with a spectrum of weapons for fighting. For increased endurance, the SACT section is also equipped with a semi-autonomous resupply UGV.

In order to quantify the benefits of the SACT relative to a conventional combat team a mathematical model is required. Consequently this paper introduces 
a simple additive weighting (SAW) multi-criteria decision making (MCDM) approach \cite{thakkar}.
The key to this is to produce a measure of combat team survivability in terms of six warfighting functions identified through Australian military doctrine. These warfighting functions will require models for their application, and so based upon observed characteristics of a dismounted combat team, a series of such models will be proposed. Inherent in these models is a random process accounting for the first time that the combat team experiences an event of significance, such as engagement with an enemy force. This results in these models changing dynamically to reflect the impact of an event of significance on the combat team. The adopted definition of combat team survivability will be based upon the assumption that the warfighting functions are independent, so that direct interactions between these functions will not considered.

The scope of the paper is to develop this model for assessing operational effectiveness of dismounted combat teams, and to illustrate how it can be utilised. The model itself has been constructed through discussions with military subject matter experts (SMEs) on the causal relationships between warfighting functions for dismounted combat operations. The model parameterisation used in this paper is arbitrary; in future studies direct SME input will be used to provide explicit assessments of dismounted combat team's expected performance.

As a preliminary to the model development it is necessary to introduce these warfighting functions and to discuss their associated dependencies, as well as restrictions imposed throughout this study. This is the focus of the next section.

\section{Warfighting Functions}
There are six warfighting functions which can be used to describe military characteristics within the combat space.
There is variation in the labelling of these functions between different countries, but the specification below is sufficient for this study, and has been adopted from \cite{talisman17}.
The warfighting functions are referred to as {\em Intelligence} (I), {\em Command and Control} (C), {\em Movement and Manoeuvre} (M), {\em Lethal and Non-Lethal Fires} (F), 
{\em Logistics} (L) and {\em Force Protection} (P). These definitions are somewhat self-explanitory but some explanation relative to a dismounted combat team is appropriate.
{\em Intelligence} encompasses developing an understanding of the combat field's situational awareness, enemy location and capability. {\em Command and Control} refers to a combat team's chain of command and the way in which it is coordinated. {\em Movement and Manoeuvre} refers to a combat team's agility in the combat space, while {\em Lethal and Non-Lethal Fires} refers to its firepower and countermeasure capability. {\em Logistics} refers to a combat team's supply chain and ability to sustain itself while {\em Force Protection} refers to a combat team's ability to defend itself. 

Throughout this study a caveat imposed is that  a dismounted combat team has embarked upon a known mission, in a particular vignette, and that underlying this vignette is the potential for engagement with an enemy force. The combat team will enter the theatre of operation with predetermined capabilities, such as weaponry and communications devices. 

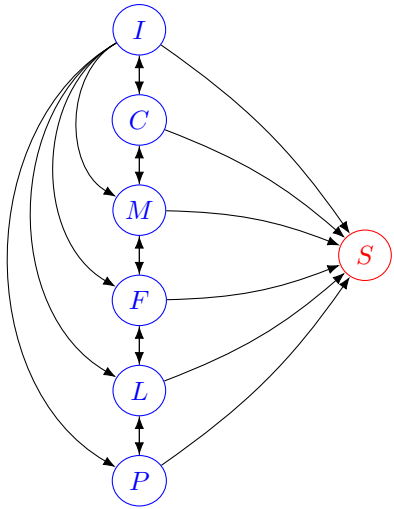
\begin{figure}[h]
\centering
\begin{tikzpicture}[scale=0.6]

\node[state, blue] (I) at (0, 12){$I$};
\node[state, blue] (C) at (0, 10){$C$};
\node[state, blue] (M) at (0, 8){$\!M\!$};
\node[state, blue](F) at (0, 6){$F$};
\node[state, blue](L) at (0, 4){$L$};
\node[state, blue](P) at (0, 2){$P$};

\node[state, red] (S) at (5, 7){$S$};
%\node[draw, dotted, inner sep=2mm , fit=(I) (C)]{};
%\node[draw, dotted, inner sep=2mm , fit=(M) (F)]{};
%\node[draw, dotted,  inner sep=2mm , fit=(L) (P)]{};

%\path (1, 7) edge (S);
%\path (1, 11) edge[bend left=20] (S);
 
\path (I) edge[bend left=10] (S);
\path (C) edge[bend left=10] (S);
\path (M) edge[bend left=10] (S);
\path (F) edge[bend right=10] (S);
\path (L) edge[bend right=10] (S);
\path (P) edge[bend right=10] (S);

\path (I) edge[bend left=-60] (M);
\path (I) edge[bend left=-60] (F);
\path (I) edge[bend left=-60] (L);
\path (I) edge[bend left=-60] (P);

%\path (1, 3) edge[bend left=-20] (S);

%\path (-1, 3) edge[bend left=40] (-1, 7);
%\path (-1, 7) edge[bend left=40] (-1, 11);
%\path (-1, 3) edge[bend left=60] (-1, 11);

%\path (1, 11) edge[bend left=60] (1, 3);
%\path (1, 11) edge[bend left=60] (1, 7);
%\path (1, 7) edge[bend left=60] (1, 3);

\path (P) edge (L);
\path (L) edge (P);
\path (L) edge (F);
\path (F) edge (L);
\path (F) edge (M);
\path (M) edge (F);
\path (M) edge (C);
\path (C) edge (M);
\path (C) edge (I);
\path (I) edge (C);

\end{tikzpicture}
\caption{A causal model illustrating some potential dependencies between metric {\em Survivability} (S) and the six warfighting functions {\em Intelligence} (I), {\em Command and Control} (C), {\em Movement and Manoeuvre} (M), {\em Lethal and Non-Lethal Fires} (F), {\em Logistics} (L) and {\em Force Protection} (P). For the purposes of this study, only the direct influences from a warfighting function node to {\em Survivability} will be considered.}
\label{fig1}
\end{figure}

There are many possible definitions of the ability of a combat team to achieve its mission success. For the purposes of this study the combat team's {\em Survivability} (S) is a suitable metric for this. {\em Survivability} can be specified in terms  of the six warfighting functions introduced above. To illustrate this Figure \ref{fig1} provides a causal model indicating some  dependencies between the warfighting functions. The causal model indicates the influences between each warfighting node and {\em Survivability}, the co-dependence between adjacent nodes and the influence that the node corrsponding to the warfighting function {\em Intelligence} has on the other five warfighting function nodes. There are potentially up to thirty influences between warfighting function nodes reflecting their co-dependence (not shown for brevity).

For analytic simplicity, one can base the performance metric {\em Survivability} on the direct influences of the warfighting functions into the {\em Survivability} node, with the proviso that the relative tradeoffs are accounted for in specific examples of warfighting functions. To illustrate this, if one was examining a SACT then its enhanced situational awareness, through increased intelligence via semi-autonomous systems, will result in a corresponding heavier demand on logistics and potential reduction in freedom of movement and manoeuvre. 
{ It is important to note that the adopted definition of {\em Survivability} implicitly assumes that
the warfighting functions are independent from a causal model perspective, and so interactions between warfighting functions are not considered directly. In further work the inclusion of interactions between warfighting functions will be examined.}

Hence the next section will quantify {\em Survivability} as a direct function of mathematical models of the six warfighting functions.

\section{Mathematical Model}
Introduce an indexing set $\U = \{I, C, M, F, L, P\}$ which indexes the warfighting functions introduced previously, and suppose that for some $u \in \U$, the function $x_u(t)$ is used to measure the warfighting function $u$ at time $t$. At this stage it will be assumed that such a model for warfighting function can be specified: it will be made explicit in the next section. In addition to this it will be assumed that these models are stochastic in nature, so that it is sensible to discuss them from a statistical perspective. Assume that $x_u(0) = \alpha_u$ is a nominal deterministic initial value, and that the function $x_u(t)$ is scaled so that its values lie in the unit interval. Hence, for example, if $x_I(0) = 1$ then this means that the combat team has perfect intelligence initially, and if $x_I(t) = 0.5$ then this implies that at time $t$ the combat team has had its intelligence reduced to 50\%. These initial values could be determined by SME judgements, or can be set to expected values based upon the warfighting function characteristics. To illustrate the latter, it may be appropriate to assume that initially the combat team has full command and control ($x_C(0) = \alpha_C = 1$) and complete logistics ($x_L(0) = \alpha_L = 1$).

Next define $\lambda_u$ to be the relative importance of the warfighting function $u \in \U$. These are also subjective choices and can, for example, also be based upon SME input to rank importance. It is assumed that each such parameter is non-negative, but must be on a consistent scale (for example percentage rankings or probabilities). They do not necessarily have to sum to unity but can also be identical in value. To illustrate this, if each warfighting function was considered equally important then one would select $\lambda_u = 1$ for all $u \in \U$.
Similarly, if one considered {\em Intelligence} and {\em Command and Control} to have higher priority than the other warfighting functions then a possible choice would be $\lambda_I = \lambda_C = 0.9$, with $\lambda_u < 0.9$ for $u \in \U - \{I, C\}$.

The two vectors $(\lambda_I, \lambda_C, \lambda_M, \lambda_F, \lambda_L, \lambda_P)$ and $(x_I(t), x_C(t), x_M(t), x_F(t), x_L(t), x_P(t))$, which are 6-dimensional vectors in the space of bounded integrable functions, capture all the essential information about the combat team at time $t$. Hence, in order to produce a SAW model, one can apply an inner product  \cite{billingsley} of these two to construct the function
\begin{equation}
Z(t) = \frac{ \sum_{u \in \U} \lambda_u x_u (t)}{\sum_{u \in \U} \lambda_u},
\label{utility1}
\end{equation}
where the vector consisting of warfighting function importance has been normalised by the sum of all its entries. 
Then by construction $Z(t) \in [0, 1]$ such that if $x_u(t) = 0$ for all $u \in \U$ then $Z(t) = 0$, and similarly if each $x_u(t) = 1$, for all $u \in \U$,  then $Z(t)$ will also be unity.
Thus \eqref{utility1} can be used as an overall measure of the combat team effectiveness since it combines the warfighting functions into a linear form and allows their relative importance to be incorporated into the measure. Additionally \eqref{utility1} can be interpreted as combat team {\em Survivability} in the causal model of Figure \ref{fig1}.

It is important to note that SME input will be critical in specifying suitable parameters and warfighting function forms for application in \eqref{utility1} to assess operational effectiveness. These choices will also be dependent on the underlying vignette and combat team operation. In the examples investigated in a subsequent section an arbitrary choice will be applied for parameterisation to illustrate the results in a somewhat general setting.

Given the complexity of deriving statistical distributions for \eqref{utility1} one can instead examine its mean, and so a practical performance metric is given by
\begin{equation}
\mean Z(t) = \frac{ \sum_{u \in U} \lambda_u \mean \left(x_u (t)\right)}{\sum_{u \in U} \lambda_u}.
\label{utility2}
\end{equation}

Expression \eqref{utility2} will therefore be utilised to undertake a comparison between the operational effectiveness of 
 two different dismounted combat teams. However, prior to this it is necessary to examine potential statistical models for warfighting functions.

\section{Models for Warfighting Functions}
A series of generic models of warfighting functions will be proposed in this section, which will be developed on the basis of an understanding of how one expects the warfighting functions to evolve over time for a dismounted combat team. 
A {\em sine qua non} in this analysis in this paper is that the models of warfighting functions remain at their initial conditions until an event takes place in the combat space. 
An event is understood to be an occurrence which has an impact on the combat team's mission, such as detection of, or engagement with, an enemy force. From a dismounted combat team perspective, the warfighting functions {\em Intelligence} and {\em Command and Control} can be assumed to remain at their initial levels until the event time. Thereafter, one would expect that the team's situational awareness would increase, while there may be a decrease in the team's command and control if the event is catastrophic. Similarly, the warfighting function {\em Movement and Manoeuvre} may be significantly reduced after the event. If the latter involves engagement with an enemy, then the warfighting function  {\em Lethal and Non-Lethal Fires} capability could be reduced, and would continue to decrease until either supplies were exhausted or the event ended. The warfighting function {\em Logistics} would remain at a fixed level, but after the event may decrease until the mission has been completed. Finally, the warfighting function {\em Force Protection} may drop significantly after the event occurred, especially if the combat team experienced a significant reduction in soldiers after an engagement. 

In view of these considerations, four types of models for warfighting functions will be proposed. These models have not been produced on the basis of elicited data but have been selected to reflect the observations discussed above.

\begin{figure}[ht]
\centering
\begin{tikzpicture}[scale=0.4] 
 \draw[thick,->] (0,0)--(8,0) node[right] {$x$}; 
    \draw[thick, ->] (0, 0)--(0, 4) node[above] {$y$};
     \draw[thin, -] (0, 2)--(3, 2);
    % \draw[thin, ->] (3, 1)--(7, 1);
    \draw[-latex] (3, 1)--(7, 1);
     \node[below] at (3,0) {$\tau$};
     \draw[thin, -] (3, -0.1)--(3, 0.1);
    \node[left] at (0,2) {$\alpha_u$};
\end{tikzpicture}
\qquad 
\begin{tikzpicture}[scale=0.4]
  \draw[thick,->] (0,0)--(8,0) node[right] {$x$}; 
    \draw[thick, ->] (0, 0)--(0, 4) node[above] {$y$};
     \draw[thin, -] (0, 2)--(3, 2);
     \draw [black] (3,1) to[out=0,in=180] (7,0.1) ; 
     \node[below] at (3,0) {$\tau$};
      \draw[thin, -] (3, -0.1)--(3, 0.1);
    \node[left] at (0,2) {$\alpha_u$};
\end{tikzpicture}
\\
\qquad
\hspace{-0.8cm}
\begin{tikzpicture}[scale=0.4]
 \draw[thick,->] (0,0)--(8,0) node[right] {$x$}; 
    \draw[thick, ->] (0, 0)--(0, 4) node[above] {$y$};
     \draw[thin, -] (0, 2)--(3, 2);
     \draw [black] (3,2) to[out=0,in=180] (7,0.1) ; 
     \node[below] at (3,0) {$\tau$};
      \draw[thin, -] (3, -0.1)--(3, 0.1);
    \node[left] at (0,2) {$\alpha_u$};
\end{tikzpicture}
\qquad
\begin{tikzpicture}[scale=0.4]
  \draw[thick,->] (0,0)--(8,0) node[right] {$x$}; 
    \draw[thick, ->] (0, 0)--(0, 4) node[above] {$y$};
     \draw[thin, -] (0, 2)--(3, 2);
     \draw [black] (3,2) to[out=0,in=180] (7,3) ; 
     \node[below] at (3,0) {$\tau$};
      \draw[thin, -] (3, -0.1)--(3, 0.1);
    \node[left] at (0,2) {$\alpha_u$};
\end{tikzpicture}
\caption{Four examples of potential models of warfighting functions, where $\tau$ indicates the time of engagement. The top left plot corresponds to the case where the warfighting function drops to a fixed level. The top right plot is for the case where the warfighting function drops as for the previous case, but then decreases with time from the point of engagement. The bottom left subplot is for the situation where the warfighting function slowly decreases with time after the engagement time, while the bottom right shows the scenario where the warfighting  function increases from the point of engagement.}
\label{fig2}
\end{figure}
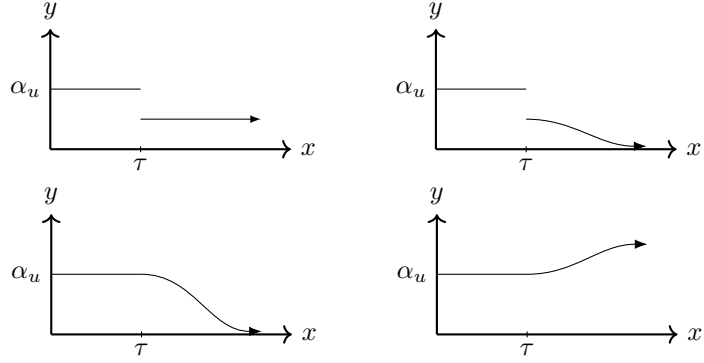

Suppose that the aforementioned event occurs at some random time $\tau$, which is assumed to be non-negative. Figure \ref{fig2} provides four examples of potential models for warfighting functions, based upon the  preceding considerations.

The top left example in Figure \ref{fig2} shows the situation where the model of warfighting function decreases to a constant value after the engagement time $\tau$. For this case
\begin{equation}
x_u(t) = \begin{cases} 
      \alpha_u & t < \tau \\
      \alpha_u R & t\geq \tau,
   \end{cases}
   \label{type1}
\end{equation}
where $R$ is a realization of a uniformly distributed random variable on the unit interval. The reason for using this random variable is to account for the fact that one cannot know {\em a priori} the level to which the warfighting function will drop at the engagement time.

Observe that \eqref{type1} can be expressed in the form $x_u(t) = \alpha_u \I[t < \tau] + \alpha_u R \I[t \geq \tau]$ where the indicator function $\I[A]$ takes the value unity if and only if condition $A$ is true, and is zero otherwise. Since it is reasonable to assume that $R$ and $\tau$ are independent, the required mean, for application in \eqref{utility2} is given by
$\mean\left(x_u(t)\right) = \alpha_u \prob(t < \tau) + 0.5 \alpha_u \prob(t \geq \tau)$, since the mean of the relevant uniform random variable is 0.5. If events in the combat space were to be modelled by a Poisson process, then the time to the first event would have an exponential distribution. Hence, under the assumption that $\tau$ has an exponential distribution with parameter $\kappa$, one can show that
\begin{equation}
\mean\left(x_u(t)\right) =0.5 \alpha_u \left[ 1 + e^{-\kappa t}\right]. \label{meanval1}
\end{equation}

A second example of a model for warfighting functions is given in the top right hand side of Figure \ref{fig2}, which illustrates the case where it decreases after the engagement time, followed by a continuous decrease with time until the function limits to zero. For a specific example of this, consider the function defined by
\begin{equation}
x_u(t) = \begin{cases} 
      \alpha_u & t <  \tau \\
      \alpha_u R e^{-\mu_u(t-\tau)}& t\geq \tau,
   \end{cases}
   \label{type2}
\end{equation}
where $R$ is again a realization of a uniformly distributed random variable on the unit interval, and $\mu_u$ is a nonnegative constant controlling the rate of decrease of the warfighting function. By expressing \eqref{type2} in terms of indicator functions, and then taking expectations, one can show that
\begin{equation}
\mean\left(x_u(t)\right) = \alpha_u e^{-\kappa t} + 0.5\alpha_u \frac{\kappa}{\kappa-\mu_u}\left(e^{-\mu_u t} - e^{-\kappa t}\right), \label{meanval2}
\end{equation}
where it has been assumed that the time $\tau$ has an exponential distribution with parameter $\kappa$.

If there was not a sudden drop in the warfighting function, but nonetheless a gradual decline after the engagement time, then this is illustrated in the bottom left subplot of Figure \ref{fig2}.  A specific example of this is given by
\begin{equation}
x_u(t) = \begin{cases} 
      \alpha_u & t < \tau \\
      \alpha_u e^{-\mu_u(t-\tau)} & t\geq \tau,
   \end{cases}
   \label{type3}
\end{equation}
where the constant $\mu_u$ regulates the rate of decrease of the warfighting function. It follows from \eqref{meanval2} that the mean of \eqref{type3} is given by
\begin{equation}
\mean\left(x_u(t)\right) = \alpha_u e^{-\kappa t} + \alpha_u \frac{\kappa}{\kappa-\mu_u}\left(e^{-\mu_u t} - e^{-\kappa t}\right), \label{meanval3}
\end{equation}
again under the same distributional assumption imposed upon $\tau$.

The bottom right subplot in Figure \ref{fig2} provides an illustration of the case where after the engagement time the warfighting function increases. As an example, this may correspond to the situation where, following an event in the combat space,  the combat team's situational awareness improves.

An explicit example for this is given by
\begin{equation}
x_u(t) = \begin{cases} 
      \alpha_u & t < \tau \\
      \beta_u-(\beta_u-\alpha_u)  e^{-\mu_u(t-\tau)} & t\geq \tau,
   \end{cases}
   \label{type4}
\end{equation}
where $\mu_u$ is a constant controlling the rate of increase and $\beta_u$ is the limit to which the situational awareness can increase ($\alpha_u < \beta_u \leq 1$).
Using a similar calculation as in the previous cases it can be shown that
\begin{equation}
\mean\left(x_u(t)\right) = \beta_u + \frac{ \beta_u - \alpha_u }{\kappa - \mu_u}\left( \mu_u e^{-\kappa t} - \kappa e^{-\mu_u t}\right), \label{meanval4}
\end{equation}
again with the assumption that $\tau$ has an exponential distribution with parameter $\kappa$.

The four means of models of warfighting functions \eqref{meanval1}, \eqref{meanval2}, \eqref{meanval3} and \eqref{meanval4} can now be applied to \eqref{utility2}, together with choices for $\lambda_u$, once the models \eqref{type1}, \eqref{type2}, \eqref{type3} and \eqref{type4} are mapped to particular warfighting functions. The next section will provide some examples of this to illustrate the utility of the approach.

\section{Particular Examples}
To illustrate the way in which the results can be applied consider the case where two combat teams, denoted Team 1 and Team 2 respectively, are facing a particular mission. Throughout the following it will be assumed that Team 1 has attributes of the SACT introduced previously, while Team 2 is a conventional combat team. The model parameterisation is arbitrary for the purposes of this study. It will be assumed that the time to an occurrence of an engagement has an exponential distribution with rate $\kappa = 0.1$. The warfighting functions are modelled as follows: {\em Intelligence} is modelled through \eqref{type4};
{\em Command and Control} and {\em Movement and Manoeuvre} through \eqref{type1}; {\em Lethal and Non-Lethal Fires} and {\em Force Protection} through \eqref{type2}; and {\em Logistics} through \eqref{type3}. Specific model parameters are documented in Table \ref{table1}.
For the case of {\em Intelligence}, the model for Team 1 has been selected so that it begins at a larger initial value than Team 2, and such that it increases faster to an upper limit of 1 than the model for Team 2, which can only increase to a maximum of 0.8. This is to reflect the fact that Team 1 is assumed to have superior sensing and intelligence gathering capabilities. Similarly, the parameterisation for {\em Command and Control} is such that it is assumed that Team 1 has superior capability, but in the case of an event, both teams could experience a sudden decline in this warfighting function capability.
In the case of {\em Movement and Manouuvre} it is assumed that Team 2 will have superiority initially, while both teams will experience a sudden drop in this warfighting function after an engagement. Additionally, Team 1 is assumed to have somewhat reduced mobility due to its increased technological load and resupply necessity. Since Team 1 will have superior weaponry it is supposed that initially its {\em Lethal and Non-Lethal Fires} model value is higher, and its model tail will drop off faster than that of Team 1. However, the parameterisation penalises Team 1 in terms of its {\em Logistics} warfighting function, since Team 1 will have a greater dependency on resupply. By contrast Team 2 will be assumed to carry all their resources. Finally, the warfighting function {\em Force Protection} is assumed to be initially the same for both teams, but will reduce faster for Team 2 after an engagement, since Team 1 can provide force protection through its semi-autonomous robotic systems.

\begin{figure}[t]
\begin{center}
\includegraphics[width=8cm]{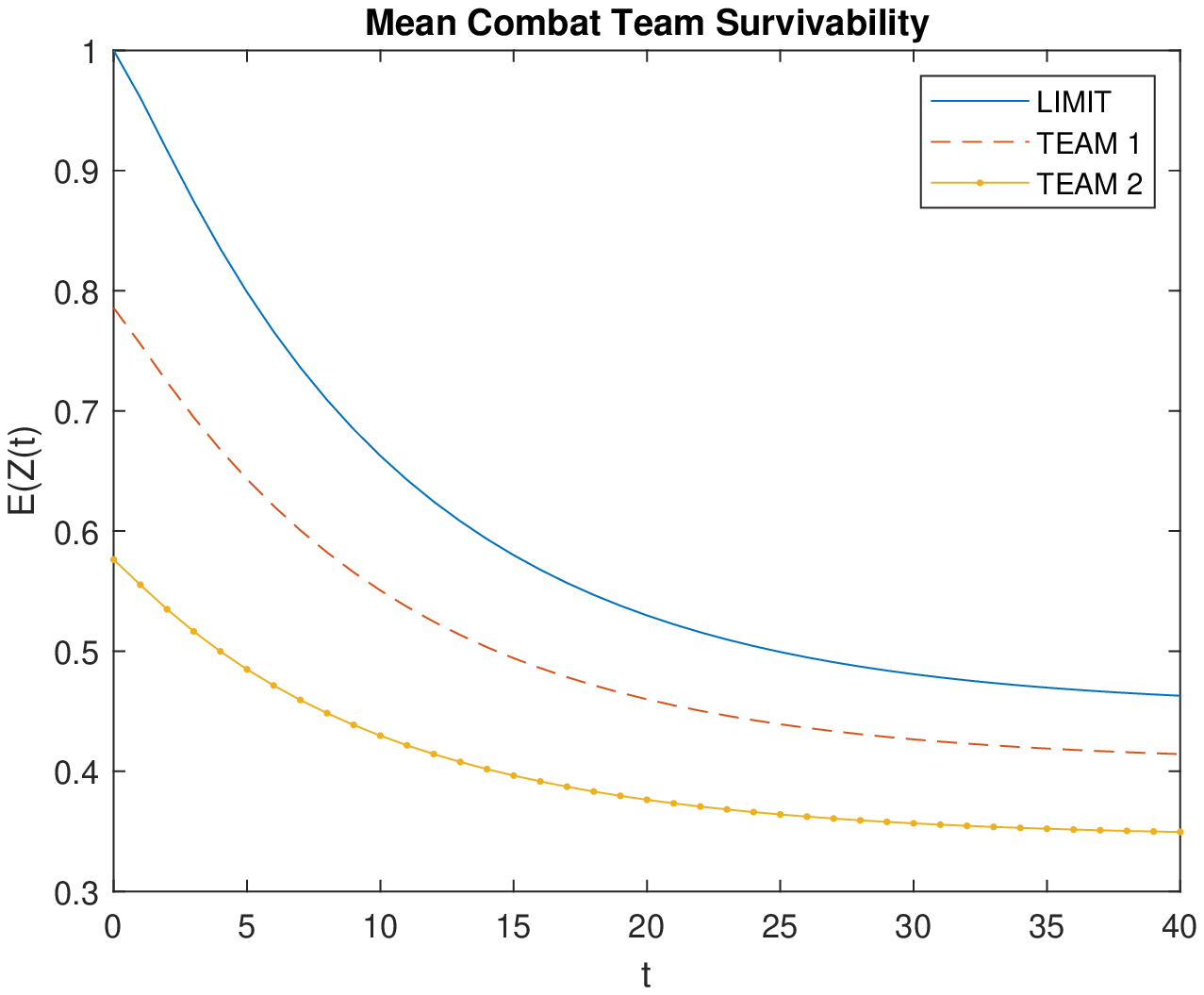} 
\includegraphics[width=8cm]{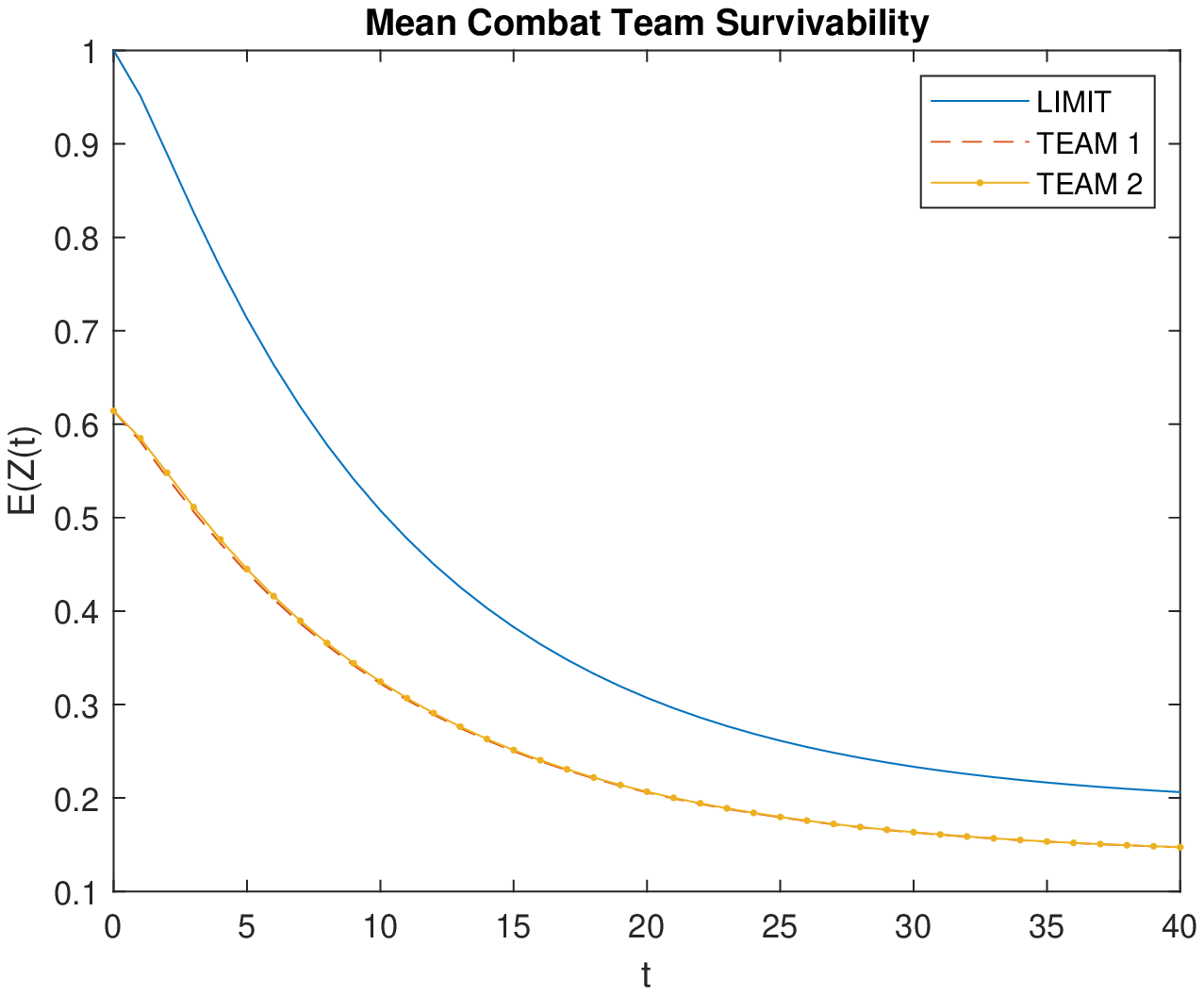}\\
\includegraphics[width=8cm]{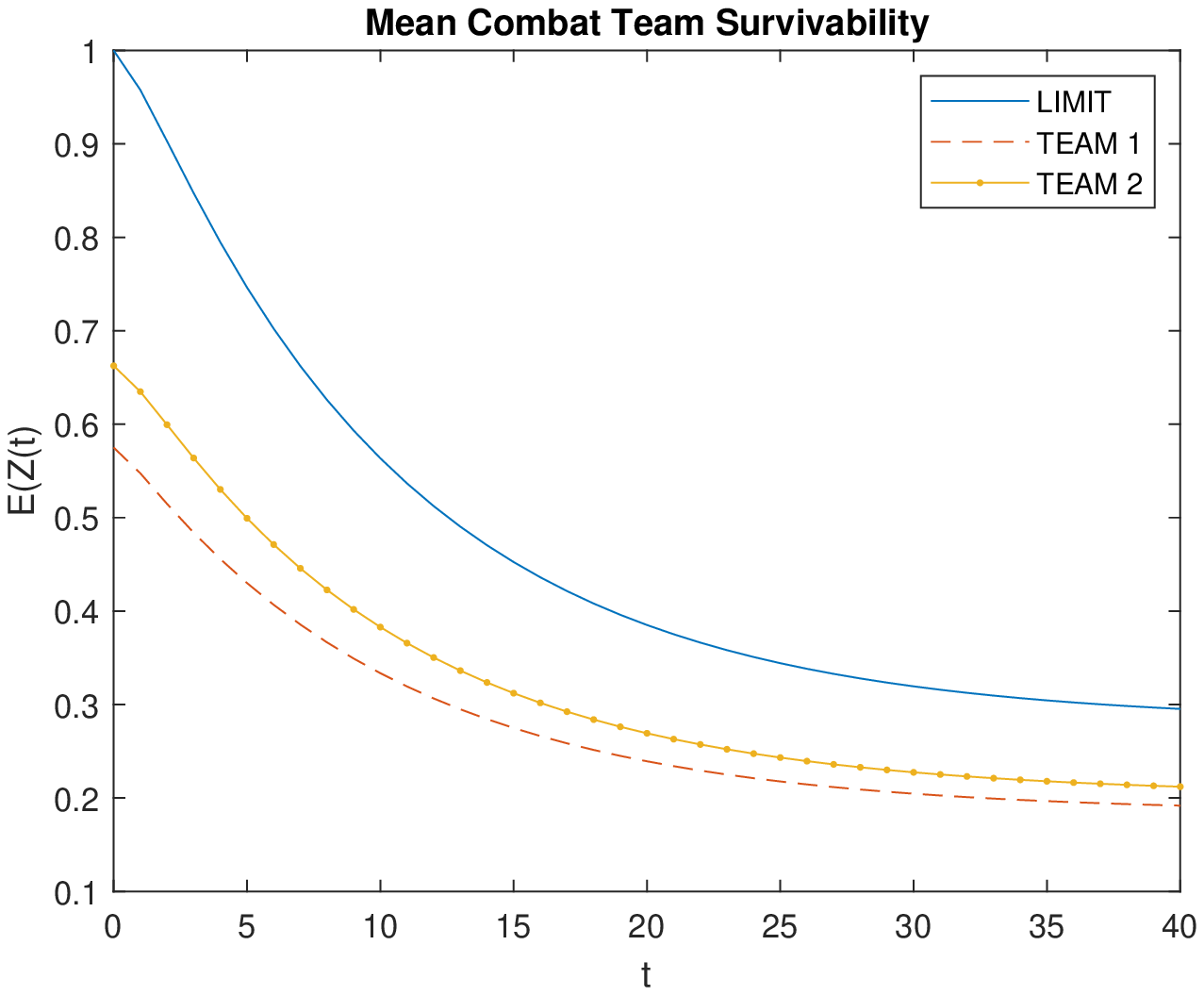} 
\includegraphics[width=8cm]{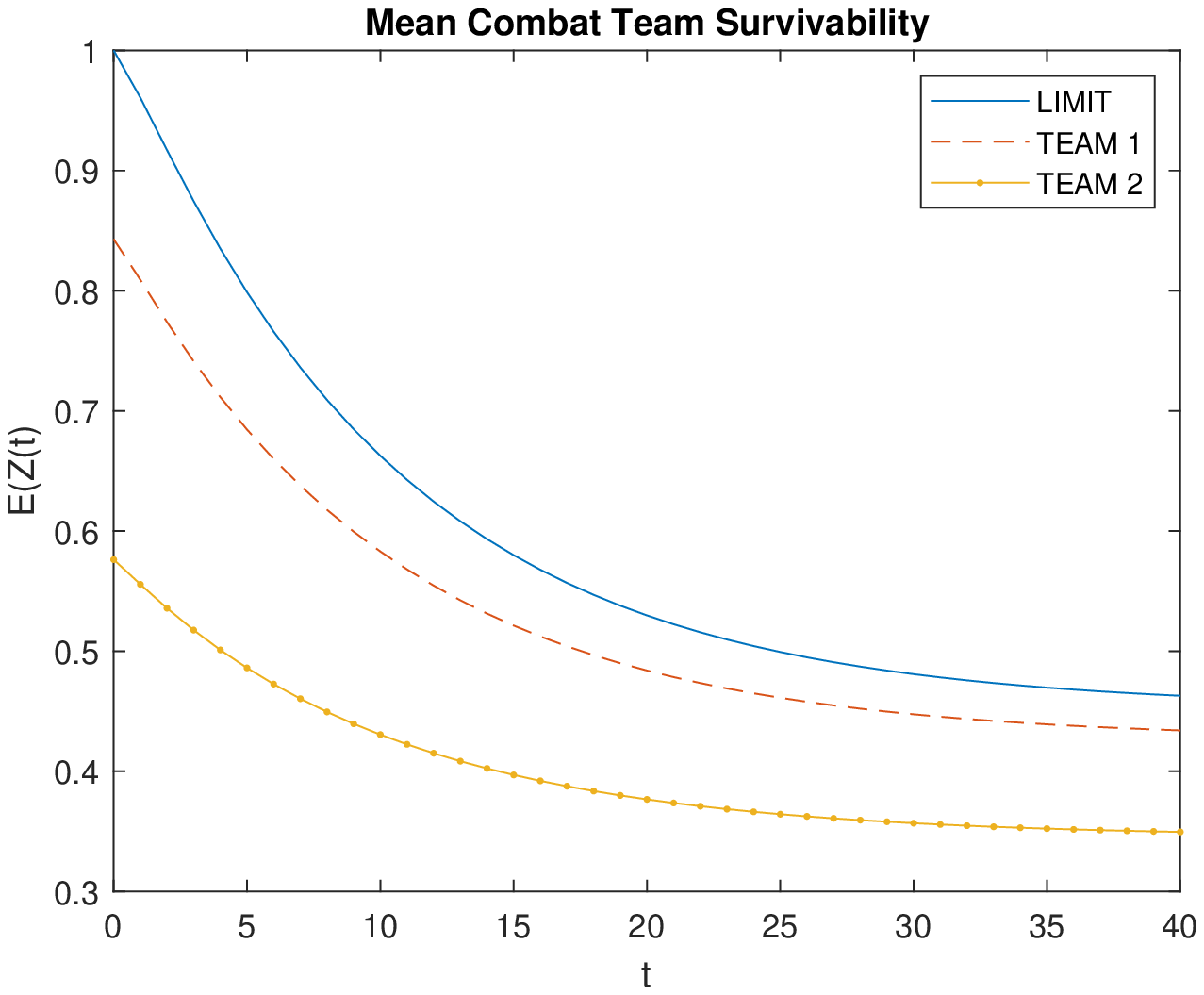}\\
\caption{Four examples of mean combat team survivability, under the parameterisation in Table \ref{table1}, and with different warfighting function rankings.}
\label{fig3}
\end{center}
\end{figure}

Figure \ref{fig3} provides a set of four examples, where the mean combat team survivability \eqref{utility2} is plotted, based upon the parameterisation in Table \ref{table1}, for both combat teams. The difference between each subplot is the choice made for the warfighting function importance ranking.

In the top left subplot the rankings have been selected to be $\lambda_I = 6, \lambda_C = 4, \lambda_M = 3, \lambda_F = 5, \lambda_L = 1, \lambda_P = 2$. Hence under this parameterisation the warfighting function {\em Intelligence} is ranked as the most important, while {\em Logistics} has been ranked the least important. The plot also shows an upper bound, denoted LIMIT, which has been based upon the same ranking applied to \eqref{utility2} but with all parameters of the warfighting function model set to their potential maxima. To clarify this, it means that all initial values of models are set to unity and rates of increase of exponential functions defined in the models of warfighting functions are set to unity. Based upon the plot in the top left subplot of Figure \ref{fig3} one concludes that under this parameterisation Team 1 has better average survivability than Team 2.

The second example can be viewed in the top right subplot of Figure \ref{fig3}, and corresponds to ranking given by
$\lambda_I = 1, \lambda_C = 2, \lambda_M = 4, \lambda_F = 3, \lambda_L = 6, \lambda_P = 5$. Here one observes that there is no difference between the two combat teams' expected survivability. Under this ranking the warfighting functions {\em Intelligence} and {\em Command and Control} have been ranked least important, while {\em Force Protection} and {\em Logistics} have been assigned more importance.

From a tactical perspective, the combat team's situational awareness and coordination are critical to their survival and mission success. Hence the ranking applied in the first example is more likely to reflect the reality in an operational setting than that imposed in the second example. Therefore from an operational effectiveness perspective, the results in the top left subplot are more indicative of expected performance under a reasonable warfighting importance ranking.

To provide further insight into this, consider the bottom left subplot in Figure \ref{fig3}. In this example the warfighting function ranking has been done somewhat differently. Here the rankings are such that $\lambda_I = \lambda_C = \lambda_F = \lambda_P = 1$, while $\lambda_M = \lambda_L = 6$.
Hence this case discounts warfighting functions where Team 1 would be superior, while the two warfighting functions where Team 2 is superior are ranked six times more important. The figure shows that Team 2 has better performance, but this is based upon a ranking that is not a reflection of reality.
The final example can be found in the bottom right subplot of Figure \ref{fig3}, where the warfighting functions have been ranked 
$\lambda_I = \lambda_C = \lambda_F = 6$ and $\lambda_M = \lambda_L = \lambda_P = 1$. Hence Team 1's strengths have been given six times more importance than its weaknesses. As can be observed from the subplot, Team 1 has significantly better expected performance than Team 2 under this ranking.

Based upon the discussion above it can be observed that the ranking of warfighting function importance has a significant effect on the results of the assessment of operational effectiveness of dismounted combat teams. However, the analysis provided in this paper results in the conclusion that under the parameterisation of warfighting functions in Table \ref{table1}, together with a reasonable ranking to reflect the importance of warfighting functions, one can conclude that Team 1 will have better expected survivability than Team 2.

\begin{table}
\centering
\begin{tabular}{ |l|l|l| }
\hline
\multicolumn{3}{ |c| }{Model Parameters} \\
\hline
\multirow{6}{*}{Team 1}   
& I & \eqref{type4}: $\alpha_I  = 0.9, \mu_I = 2, \beta_I = 1$ \\
& C & \eqref{type1}: $\alpha_C = 0.9$ \\ 
& M & \eqref{type1}: $\alpha_M = 0.5$ \\    
 & F & \eqref{type2}: $\alpha_F  = 0.9, \mu_F = 2$ \\
 & L & \eqref{type3}: $\alpha_L = 0.5, \mu_L = 2$ \\
 & P & \eqref{type2}: $\alpha_P = 0.5, \mu_P = 0.5$ \\
 \hline
\multirow{6}{*}{Team 2} 
& I & \eqref{type4}:  $\alpha_I = 0.5, \mu_I = 0.5, \beta_I = 0.8$ \\ 
& C & \eqref{type1}: $\alpha_C = 0.7$ \\ 
& M & \eqref{type1}: $\alpha_M = 0.7$ \\ 
& F & \eqref{type2}: $\alpha_F = 0.5, \mu_F = 0.5$ \\ 
& L & \eqref{type3}: $\alpha_L = 0.7, \mu_L = 1$ \\
 & P & \eqref{type2}: $\alpha_P = 0.5, \mu_P = 2$ \\
\hline
\end{tabular}
\caption{Model parameterisation adopted for the study. For each combat team type, and for each warfighting function, the model type is indicated, together with parameters adopted for each case.}
\label{table1}
\end{table}

\section{Conclusions}
The purpose of this paper has been to introduce a framework in which the operational effectiveness of a dismounted combat team could be assessed. Through the identification of six warfighting functions, a metric {based upon a SAW model} has been introduced which has been based upon warfighting function importance ranking and models for each such warfighting function. The latter models have been based upon an understanding of how warfighting functions are likely to evolve over time, and in the presence of an event of significance. By utilising the mean combat team survivability metric it was shown how two dismounted combat teams could be compared in terms of performance. Under the adopted model parameterisation it was demonstrated that expected performance is sensitive to the warfighting function importance ranking. Nonetheless, it was possible to argue that the dismounted combat team with enhanced functionality (the SACT) had better expected survivability than a conventional combat team, with the assumed model parameterisation.

Further work will consider the interaction of warfighting functions more directly and their impact on the combat team survivability metric. In addition to this it will be of significant importance to validate the choices for warfighting function models, as well as model parameterisation and warfighting function ranking. This is a future exercise planned to be undertaken with military SMEs at DSTG.

%%%%%%%%%%%%%%%%%%%%%%%%%%%%%%%%

%\clearpage

\end{document}